\documentclass[runningheads]{svmult}

\usepackage{makeidx}   
\usepackage{graphicx}  
\usepackage{subeqnar}  
\usepackage{multicol}  
\usepackage{physprbb}  
\makeindex             

\begin{document}
\title*{Abundances as Tracers of the Formation and
\protect\newline Evolution of (Dwarf) Galaxies}
\toctitle{Abundances as Tracers of the Formation and
\protect\newline Evolution of (Dwarf) Galaxies}
%
%
\titlerunning{Abundances as tracers}
%
\author{Eline Tolstoy}
\authorrunning{Tolstoy}
%
%
\institute{Kapteyn Institute, University of Groningen, 
9700AV Groningen, the Netherlands}

\maketitle              

\begin{abstract}
This aims to be an overview of what detailed observations of
individual stars in nearby dwarf galaxies may teach us about galaxy
evolution. This includes some early results from the DART (Dwarf
Abundances and Radial velocity Team) Large Programme at ESO. This
project has used 2.2m/WFI and VLT/FLAMES to obtain spectra of large
samples of individual stars in nearby dwarf spheroidal galaxies and
determine accurate abundances {\it and} kinematics.  These results can
be used to trace the formation and evolution of nearby galaxies from
the earliest times to the present.
\end{abstract}

\section{Introduction}

Dwarf galaxies are the most numerous type of galaxy we know of and
they are commonly assumed to be if not the actual building blocks of
larger galaxies then they most closely resemble them.  The Local Group
contains $\sim$36 dwarf galaxies out of a total of $\sim$42 members
covering a large range of properties \cite{mateo98}, and including
more than one example of most if not all the known classes of dwarf
galaxy. There are nucleated dwarfs (e.g., NGC205, NGC185); extremely
low surface brightness dwarfs (e.g., Sextans, Ursa Minor); interacting
dwarfs (e.g., Sagittarius); star bursting dwarfs (e.g., IC10, Sextans
A); isolated dwarfs (e.g., Tucana, Cetus). They fall predominantly
into two classes - those with gas which are still forming stars and
those which appear not to have gas and are not presently forming
stars.

The abundance patterns of individual stars of different ages and 
environments enable us to unlock the evolutionary history
of galaxies. Many physical characteristics of a
galaxy may change over time, such as shape and colour, however the metal
content and abundance ratios of stellar atmospheres are not so easy to
tamper with. Stars retain the chemical imprint of the interstellar gas
out of which they formed, and metals can only increase with time. This
method to study galaxy evolution has been elegantly named {\it
Chemical Tagging} \cite{freebh02}.


There have been a number of detailed abundance studies of stars in
nearby galaxies which cast ever more serious doubt on the premise that
the galaxies we see today are in any way related to galactic building
blocks (e.g., \cite{sh03}, \cite{et03}, \cite{v04}).  The
[$\alpha$/Fe] ratios of stars in dwarf 
spheroidal (dSph) galaxies are generally
lower than similar metallicity Galactic stars.  There is marginal
overlap in the [$\alpha$/Fe] ratios between dSph stars and Galactic
halo stars but this similarity does not extend to other element ratios
where, for example, a significant over abundance in [Ba/Y] is 
typically observed
in dSph stars compared to Galactic stars (see Venn, this
volume).  The stars in larger galaxies, such as the LMC and 
Sagittarius are also chemically distinct from the
majority of the Galactic stars (see Hill, this volume; Bonifacio, this
volume; McWilliam, this volume).  This makes a merging hypothesis
difficult to explain any component of our Galaxy, unless the merging
were to occur predominantly at very early times.  These observations
can thus be interpretated in two ways - either to say that dwarf
galaxies are not building blocks of larger galaxies, which begs the
questions - what are they then?  and how do they avoid too much
merging with larger galaxies? or we can say that perhaps the whole
idea of heirarchical structure formation as it currently stands needs
some serious revision because dwarf galaxies really don't fit the
picture. Both interpretations have merit and neither can be ruled out.

\section{Initial Results from DART: Sculptor Dwarf Spheroidal}

The DART large programme at ESO made v$_{hel}$ and [Fe/H] measurements
from FLAMES spectroscopy of 401 red giant branch (RGB) stars in the
Sculptor (Scl) dSph \cite{et04}.  The relatively high signal/noise,
S/N ($\approx$ 10-20 per pixel) resulted in both accurate metallicites
($\approx$ 0.1 dex from internal errors) and radial velocities
($\approx \pm$2 km/s).  This is the first time that a large sample of
accurate velocities {\it and} metallicities have been measured in a
dwarf galaxy.

Scl is a close companion of the Milky Way, at a distance of 72 $\pm
5$~kpc \cite{kudem77}, with a low total (dynamical) mass, (1.4 $\pm
0.6) \times 10^7 M_\odot$ \cite{quel95}, and modest luminosity, M$_V =
-10.7 \pm 0.5$, and central surface brightness, $\Sigma_{0,V} = 23.5
\pm 0.5$ mag/arcsec$^2$ \cite{irhat95} with no HI gas
\cite{sclh1}. CMD analysis, including the oldest Main Sequence
turnoffs, has determined that this galaxy is predominantly old and
that the entire star formation history can have lasted only a few Gyr
\cite{monk99}.

\begin{figure}[!ht]
\begin{center}
\includegraphics[angle=270,width=0.95\textwidth]{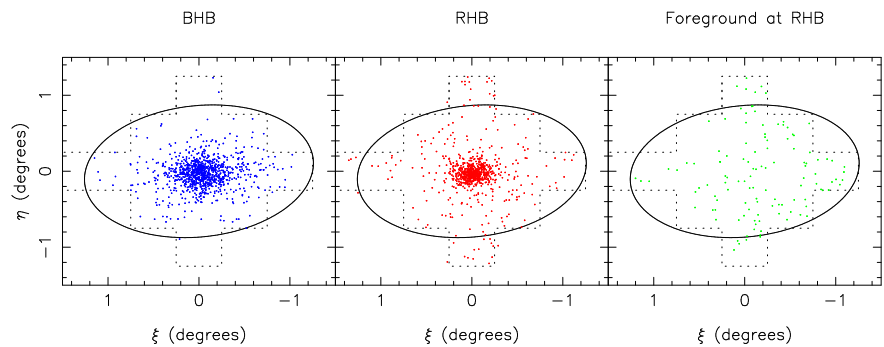}
\end{center}
\caption[]{
The distribution of Horizontal Branch stars from WFI imaging of the
Scl dSph showing the different spatial distributions of BHB and RHB as
selected from a M$_{v}$, V$-$I Colour-Magnitude Diagram \cite{et04}.
Also shown, to illustrate the foreground contamination in the RHB
distribution, are a CMD-selected sample of foreground stars to match
the RHB contamination density.  The ellipse is the tidal radius of
Scl.
}
\label{eps1}
\vskip -0.5cm
\end{figure}

{\bf Imaging:} Previous studies already suggested that the spatial
distribution of the Horizontal Branch stars in Scl shows signs of a
gradient, with the red horizontal branch stars (RHB) being more
centrally concentrated than the blue horizontal branch (BHB) stars
\cite{hk99}, \cite{maj99}. The DART WFI imaging data extends beyond
the nominal tidal radius and with an average 5-$\sigma$ limiting
magnitude of V$=$23.5 and I$=$22.5 also probes well below the
Horizontal Branch.  This has enabled us to unequivocally demonstrate
that the BHB and RHB stars have markedly different spatial
distributions (see Fig.~1).

The different spatial occupancy of the two populations, taking into
account the foreground contamination present in the RHB sample, is
striking and provides strong evidence that we are seeing two distinct
components.  The characteristics of the BHB and RHB are also
consistent with different ages (e.g., $\leq$2~Gyr), or different
metallicities ($\Delta$[Fe/H]$\sim$0.7 dex), from theoretical
modelling of globular cluster Horizontal Branches \cite{lee01}.


\begin{figure}[!ht]
\begin{center}
\includegraphics[width=.95\textwidth]{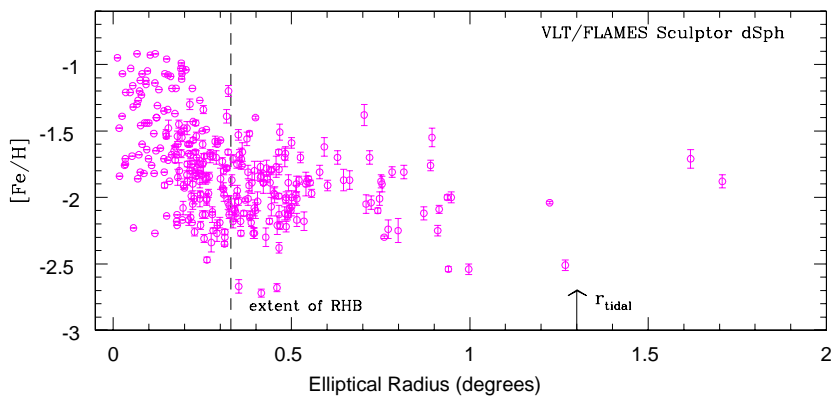}
\vskip -6cm
\end{center}
\caption[]{
VLT/FLAMES [Fe/H] measurements for 308 potential members of Sculptor
dSph versus elliptical radius. Also marked, is the extent of the RHB
distribution (as a dashed line) and the tidal radius, r$_{tidal}$.
}
\label{eps3}
\end{figure}

{\bf Spectroscopy}: The WFI images were used to select samples of
stars on the RGB in Scl to take spectra in the Ca~II triplet region
with VLT/FLAMES. This resulted in radial velocity measurements and
metallicity estimates for more than 400 stars in Scl over the fields
outlined in Fig.~1, of which 300 have a high membership probability.

In Fig.~2 we show the distribution of [Fe/H] as a function of
elliptical radius (the equivalent distance along the semi-major axis
from the centre of Scl) for those RGB stars which were determined to
have a high probability of membership. A well-defined metallicity
gradient is apparent with a similar scale size to the RHB versus BHB
spatial distributions.

In the central region of Sculptor we have high resolution spectra
providing direct abundance measurements for numerous elements (Hill et
al., in prep). In Fig.~3 we show the preliminary results for the 
$\alpha$-elements (Ca, Mg \& Ti) compared with similar
observations of stars in our Galaxy.  It is clear that the overall
distribution of [$\alpha$/Fe] versus [Fe/H] in Scl does not match our
Galaxy, except for a small number of the most metal poor stars in
Scl which overlap with Galactic halo stars.

\begin{figure}[!ht]
\begin{center}
\includegraphics[width=.95\textwidth]{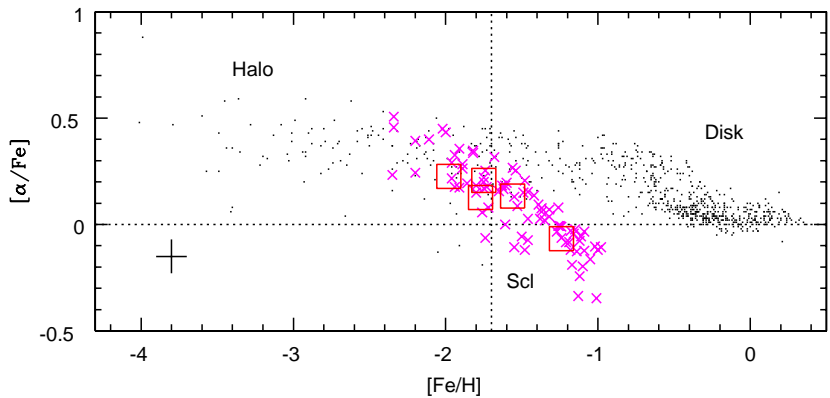}
\vskip -6cm
\end{center}
\caption[]{
The $\alpha$-abundance (average of Ca, Mg and Ti) for stars in our
Galaxy compared to those in Scl. The VLT/FLAMES high resolution
measurements of 92 members in the central field are shown as crosses
(from Hill et al., in prep).  The Galactic stars come from standard
literature sources (see \cite{v04} for references). The 5 open squares
are UVES measurements of individual stars in Scl \cite{sh03}.
}
\label{hr}
\vskip -0.5cm
\end{figure}

\vskip -2cm
\section{Two Stellar Components}

Our FLAMES results clearly 
show that Scl contains two distinct stellar
components with different spatial, kinematic and abundance properties
\cite{et04}.  There appears to be a metal-rich, $-0.9 >$ [Fe/H] $>
-1.7$, and a metal-poor, $-1.7 >$ [Fe/H] $> -2.8$ component. The
metal-rich component is more centrally concentrated than the metal
poor, and on average appears to have a lower velocity dispersion,
$\sigma_{metal-rich} = 7 \pm 1$ km/s, whereas $\sigma_{metal-poor} =
11 \pm 1$ km/s (see Battaglia, this volume).

There are indications that the presence of two populations is a common
feature of dSph galaxies.  Our preliminary analysis of HB stars,
v$_{hel}$ and [Fe/H] measurements in the other galaxies in our sample
(Fornax and Sextans dSph; Battaglia et al., in prep) also shows very
similar characteristics to Scl, especially in the most metal poor
component. Pure radial velocity studies \cite{wilk04}, \cite{kley04}
have also considered the possibility that kinematically distinct
components exist in Ursa Minor, Draco and Sextans dSph galaxies.

What mechanism could create two ancient stellar components in a small
dwarf spheroidal galaxy?  A simple possibility is that the formation
of these dSph galaxies began with an initial burst of star formation,
resulting in a stellar population with a mean [Fe/H] $\leq - 2$.
Subsequent supernovae explosions from this initial episode could have
been sufficient to cause gas (and metal) loss such that star formation
was inhibited until the remaining gas could sink deeper into the
centre \cite{mori02}.  Thus the subsequent generation(s) of stars
would inhabit a region closer to the centre of the galaxy, and have a
higher average metallicity and different kinematics.  Another possible
cause is external influences, such as minor mergers, or accretion of
additional gas.  It might also be that events surrounding the epoch of
reionisation influenced the evolution of these small galaxies
\cite{skill03} and resulted in the stripping or photoevaporation of
the outer layers of gas in the dSph, meaning that subsequent more
metal enhanced star formation occured only in the central regions.

The full abundance analysis of the FLAMES HR data (Hill et al. in
prep) will provide more details of the chemical enrichment history of
Scl. This will hopefully enable us to distinguish between two episodes
of star formation or more continuous star formation, manifested as a
gradient in velocity dispersion and metallicity from the centre of the
galaxy. Fig.~3 suggests that [$\alpha$/Fe] differs for the two
populations, such that the metal poor (presumably older) population
has high [$\alpha$/Fe] consistent with the halo of our Galaxy, and the
more metal rich population doesn't match any of the Galactic stars.
The dotted line is drawn at [Fe/H]$ = -1.7$ to show the proposed
dividing line between the two populations, although as can be seen in
Fig.~2 there is not a clear division, however the kinematics provide
clear support\cite{et04}.



\section{Dwarf Galaxies and Galaxy Formation}

It is clear from Fig.~3 that the Scl dSph does not, in the mean, have
stellar abundance properties consistent with our Galaxy, and the
evidence points to this also being the case for most other nearby
galaxies. This suggests that nearby dwarf galaxies are not the
building blocks left over from the heirarchical formation of galaxies
like our own.  If we wish to retain a heirarchical formalism to
explain the formation and evolution of our Galaxy then a mechanism has
to be found by which those objects which did merge to form our Galaxy
evolved differently from the similar mass (dwarf) galaxies we see
around today. It might be, for example, that the building blocks
formed and evolved much closer in to the central potential and thus
their star formation history and chemical evolution were affected such
that these processes proceeded much more rapidly than in their more
distant cousins.  Although this can be understood in qualitative
generalised terms there is no particular evidence to support this. You
would think that a difference in the abundance patterns of stars
formed close to the centre of a potential and further out would then
suggest some kind of gradient in properties, or at least that today we
might pick up an unabsorbed building block in the form of a dwarf
galaxy. So far detailed abundances have been determined for all the
dwarf galaxies in our halo (Shetrone 2004, this volume), including
Sagittarius (Bonifacio 2004, this volume), which is in the process of
merging with our Galaxy and they all show an astonishing uniformity of
abundance ratios, for [$\alpha$/Fe] as well as r- and s- process
elements like [Ba/Eu], and [Y/Ba] even if [Fe/H] varies between $-$3
and $-$0.5 dex (see Venn 2004, this volume).  We also see evidence for
similar abundance ratios in young massive stars in more distant higher
mass dwarf irregular galaxies (see Kaufer 2004, this volume) and in
RGB stars in the Magellanic Clouds (Hill 2004, this volume).  This
uniformity in of itself suggests a remarkably stable enrichment
process. This is even more striking considering 
that all these galaxies are so 
different in their star formation histories, but 
their mean stellar abundance patterns {\it for stars of all ages}
are very similar to each other and very different to our Galaxy.

So, although we cannot rule out the possibility that we are living at
a particular time where all recent (and future) mergers will be of
different types of objects than created the bulk of our Galaxy, it
might be wise to start to consider some variations on the standard
scenario. It is possible to argue that the most metal poor tail
of stars in Scl dSph overlap the properties of the Galactic halo
stars, suggesting evidence for extremely early (gas rich) merging
meaning that most of the stars we see in our Galaxy were actually
formed there, and only the metal poor stars in the halo may have been
formed in the satellites themselves.  


\vskip 0.3cm
\noindent{\bf Acknowledgments:}
I am grateful for support from a fellowship of the Royal Netherlands
Academy of Arts and Sciences, and the exceptional collaborators that
make up DART: Vanessa Hill Mike Irwin, Pascale Jablonka, Kim Venn,
Matthew Shetrone, Amina Helmi, Giuseppina Battaglia, Bruno Letarte,
Andrew Cole, Francesca Primas, Patrick Fran\c{c}ois, Nobuo Arimoto,
Andreas Kaufer, Thomas Szeifert \& Tom Abel.

%

\end{document}